\documentclass[12pt]{article}
\usepackage{graphicx}
\usepackage{amssymb}
\newlength{\extraspace}
\newlength{\extraspaces}
\newcommand{\be}{\begin{equation}
\addtolength{\abovedisplayskip}{\extraspaces}
\addtolength{\belowdisplayskip}{\extraspaces}
\addtolength{\abovedisplayshortskip}{\extraspace}
\addtolength{\belowdisplayshortskip}{\extraspace}}
\newcommand{\ee}{\end{equation}}

\newcommand{\ba}{\begin{eqnarray}
\addtolength{\abovedisplayskip}{\extraspaces}
\addtolength{\belowdisplayskip}{\extraspaces}
\addtolength{\abovedisplayshortskip}{\extraspace}
\addtolength{\belowdisplayshortskip}{\extraspace}}
\newcommand{\ea}{\end{eqnarray}}

\newcommand{\newsection}[1]{
\vspace{12mm}
\pagebreak[3]
\addtocounter{section}{1}
\setcounter{equation}{0}
\setcounter{subsection}{0}
\noindent{\bf \thesection. #1}
\nopagebreak
\medskip
\nopagebreak}

\newcounter{saveeqn}

\newcommand{\dif}{\mathrm{d}}

\begin{document}
\addtolength{\baselineskip}{1.5mm}

\thispagestyle{empty}
\begin{flushright}

\end{flushright}
\vbox{}
\vspace{2cm}

\begin{center}
{\LARGE{A rotating black lens solution in five dimensions
        }}\\[16mm]
{Yu Chen~~and~~Edward Teo}
\\[6mm]
{\it Department of Physics,
National University of Singapore, 
Singapore 119260}\\[15mm]

\end{center}
\vspace{2cm}

\centerline{\bf Abstract}
\bigskip 
\noindent 
It has recently been shown that a stationary, asymptotically flat vacuum black hole in five space-time dimensions with two commuting axial symmetries must have an event horizon with either a spherical, ring or lens-space topology. In this paper, we study the third possibility, a so-called black lens with $L(n,1)$ horizon topology. Using the inverse scattering method, we construct a black lens solution with the simplest possible rod structure, and possessing a single asymptotic angular momentum. Its properties are then analysed; in particular, it is shown that there must either be a conical singularity or a naked curvature singularity present in the space-time.


\newpage

\newsection{Introduction}

In four space-time dimensions, it is well known that stationary, asymptotically flat black holes are uniquely determined by their asymptotic mass, angular momentum and charge. However, such uniqueness results do not apply to five or higher dimensions. A counterexample is the five-dimensional rotating black ring \cite{Emparan:2001,Emparan:2006}, with event-horizon topology $S^2\times S^1$, which may in certain cases carry the same mass and angular momentum as a Myers--Perry black hole \cite{Myers:1986} with $S^3$ event-horizon topology. Thus, unlike in four dimensions, black holes in five dimensions may have non-spherical event-horizon topologies, and the allowed topologies have been classified in \cite{Cai:2001,Helfgott:2005,Galloway:2005}.

Hollands and Yazadjiev \cite{Hollands:2007,Hollands:2007b} have recently considered how a uniqueness result might be proved for black holes in five dimensions. They showed that stationary, asymptotically flat vacuum black holes with two commuting axial symmetries are uniquely determined by their mass, angular momentum and so-called rod structure \cite{Emparan:2001a,Harmark:2004}. In particular, it is the rod structure which determines the topology of the event horizon, and it was shown that there are only three possibilities: a 3-sphere $S^3$, a ring $S^2\times S^1$, or a lens space $L(p,q)$, consistent with the results of \cite{Cai:2001,Helfgott:2005,Galloway:2005}. The first two cases are just the Myers--Perry black hole and the black ring, and it is the purpose of this paper to investigate the third possibility---a so-called {\it black lens\/}.\footnote{Black holes with lens-space horizon topology have previously been considered in the literature, but these have been in non-asymptotically flat spaces. Examples include Taub-NUT space \cite{Elvang:2005} and the more general Gibbons--Hawking space \cite{Ishihara:2006}.}

Supposing that such a black lens solution exists, Hollands and Yazadjiev \cite{Hollands:2007} showed that the simplest rod structure it could take is the one depicted in Fig.~1. In this figure, $t$ is the time coordinate while $\psi$ and $\phi$ are the two axial coordinates. The orientation of each rod with respect to the coordinates $(t,\psi,\phi)$ is indicated above it. As can be seen, there is a finite time-like rod which represents the event horizon, and two semi-infinite space-like rods which are the usual asymptotic axes. The new feature lies in the finite space-like rod, which has components in {\it both\/} the $\psi$ and $\phi$ directions. If it has orientation $(0,p,q)$ for coprime integers $p$ and $q$, then it follows that the horizon has topology $L(p,q)$. Two special cases can immediately be seen: Firstly, if its orientation is $(0,0,1)$, then the horizon topology is $L(0,1)=S^2\times S^1$ which corresponds to the black ring. Secondly, if its orientation is $(0,1,0)$, then the horizon topology is $L(1,0)=S^3$ which corresponds to the Myers--Perry black hole.

Actually, it was explained in \cite{Hollands:2007} that when two space-like rods meet at a junction, the space-like components of the two orientation vectors must have determinant $\pm1$. This is the requirement for there not to be an orbifold singularity at the junction \cite{Evslin:2008}. In the case of the two space-like rods in Fig.\ 1 meeting at $z_3$, this implies that $q=1$. If we rename $p$ more suggestively as $n$, the orientation of the finite rod will be $(0,n,1)$, and thus the allowed horizon topology for the black lens is $L(n,1)$. Without loss of generality, we may take $n$ to be positive. 

Very recently, Evslin \cite{Evslin:2008} made a first attempt towards constructing an explicit black lens solution. However, he did not consider the rod structure in Fig.\ 1, but rather one with a second finite space-like rod inserted to the left of the time-like rod, with orientation $(0,1,-n)$. The presence of this extra rod means the event horizon would have the more restrictive lens-space topology $L(n^2+1,1)$. Evslin managed to construct a static metric with this rod structure; furthermore, he found that while conical and orbifold singularities could be eliminated from the space-time, there exist spherical naked curvature singularities surrounding each of the two junctions where the space-like rods meet. He went on to conjecture that these singularities could somehow be resolved, possibly by making the black lens rotate.

\begin{figure}[t]
\begin{center}
\includegraphics{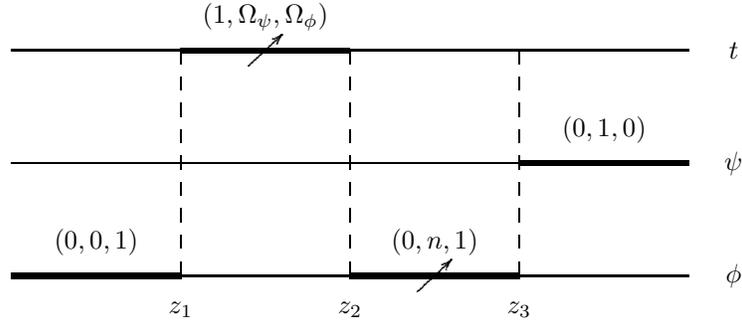}
\caption{The rod structure of the rotating black lens solution.}
\end{center}
\end{figure}

In this paper, we shall revisit the simpler rod structure of Fig.\ 1, and construct an asymptotically flat black lens solution with this rod structure. This is done using the inverse scattering method, and indeed we are able to derive a solution that possesses an asymptotic angular momentum in the $\psi$ direction. In the static limit, this solution was actually found independently in \cite{Ford:2007} and \cite{Lu:2008}, although it was not interpreted as an asymptotically flat black lens in either paper. We find that even with rotation present, the black lens either has to have a conical singularity along the finite space-like rod, or a naked singularity with spherical topology surrounding the junction $z_3$ similar to what Evslin found in his static solution. Of these two possibilities, we actually prefer the former interpretation for reasons that would be explained below.

For clarity of presentation, the static and rotating cases will be discussed separately in this paper. We begin in Sec.~2 by presenting the static black lens solution. Its properties are then analysed with particular attention paid to the global structure of the space-time, including the possible existence of conical and curvature singularities outside the horizon. The black lens with a single angular momentum is then presented and analysed in Sec.~3, emphasizing mainly the differences introduced by the rotation. Sec.~4 concludes the paper with a discussion of our results. The background and certain black-hole limits of the solution are studied in the appendices.

\newsection{Static black lens}

The space-time solution describing a static black lens was first derived in \cite{Ford:2007}, although it was not interpreted as such, being used as a stepping-stone to construct a black ring in Taub-NUT space. This solution was obtained using the inverse scattering method \cite{Belinsky:1971,Belinsky:1979,Belinski:2001}, starting from the static black ring solution with the rod structure as given in Fig.~1 with $n=0$. An anti-soliton is first removed from the point $z_3$ with a so-called BZ vector $(0,1,0)$, and then added back with a more general BZ vector $(0,1,-a)$. (For the more technical details of this procedure, the reader is referred to \cite{Pomeransky:2005,Ford:2007}.) A change in coordinates is then needed to bring the solution into an asymptotically flat form, with the rod structure exactly as in Fig.~1.

While this solution was originally derived in Weyl coordinates, it turns out to have a simpler form in C-metric type coordinates familiar from the black ring case \cite{Emparan:2004,Harmark:2004}. The explicit relation between the two coordinates can be found in Appendix H of \cite{Harmark:2004}. After this transformation, the metric reads\footnote{Some minor notational differences: coordinates $(x,y)$ are used instead of $(u,v)$ in \cite{Harmark:2004}, while $a\rightarrow-a$ and $\phi\leftrightarrow\psi$ compared to \cite{Ford:2007}.}
\ba
\label{metric}
\dif s^2=-\frac{1+cy}{1+cx}\,\dif t^2
&+&\frac{2\kappa^2(1+cx)}{(1-a^2)(x-y)^2H(x,y)}\Bigg\{
\frac{H(x,y)^2}{1-c}\left(\frac{\dif x^2}{G(x)}-\frac{\dif y^2}{G(y)}\right)\cr
&&+(1-x^2)\left[(1-c-a^2(1+cy))\dif\phi-ac(1+y)\dif\psi\right]^2\cr
&&-(1-y^2)\left[(1-c-a^2(1+cx))\dif\psi-ac(1+x)\dif\phi\right]^2\Bigg\}\,,
\ea
where the functions $G$ and $H$ are defined as
\be
G(x)=(1-x^2)(1+cx)\,,\qquad
H(x,y)=(1-c)^2-a^2(1+cx)(1+cy)\,.
\ee
As in the black ring case, $\kappa$ is a scale parameter while the parameter $c$ takes the range $0<c<1$. The new parameter $a$ takes the range $-1<a<1$ to ensure the correct space-time signature. Note that the metric is invariant under the action $a\rightarrow-a$ and either $\psi\rightarrow-\psi$ or $\phi\rightarrow-\phi$. The coordinates take the range $-\infty<t<\infty$, $-1\leq x\leq 1$, $-1/c<y\leq-1$, with $\psi,\phi$ having periodicity $2\pi$ to ensure asymptotic flatness. Asymptotic infinity is located at $x,y\rightarrow-1$, while $y=-1/c$ turns out to be the location of an event horizon. The ADM mass of this space-time can be calculated to be
\be
M=\frac{3\pi\kappa^2c}{2G}\,,
\ee
where $G$ is the gravitational constant in five dimensions.
The background limit is recovered when $c\rightarrow0$, and this will be examined in more detail in Appendix A.

In the coordinates of (\ref{metric}), the three points labelled in Fig.~1 are given by $z_1=-c\kappa^2$, $z_2=c\kappa^2$ and $z_3=\kappa^2$; while the four rods, from left to right, are located at $x=-1$, $y=-1/c$, $x=1$ and $y=-1$, respectively.
It can be checked that the orientations of the $x=-1$, $y=-1/c$ and $y=-1$ rods are $(0,0,1)$, $(1,0,0)$ and $(0,1,0)$, respectively, as expected of an asymptotically flat, static black hole. The orientation of the finite space-like rod at $x=1$ is
\be
\label{orientation}
\left(0,\frac{2ac}{1-c-a^2(1+c)},1\right).
\ee
Two special cases can immediately be read off from this result. One is when $a=0$, in which the $x=1$ rod is parallel to the $x=-1$ rod. In this event, we recover the static black ring with $S^2\times S^1$ event-horizon topology. The other is when $a=\pm\sqrt{(1-c)/(1+c)}$, in which the $x=1$ rod is parallel to (indeed, joined up to) the $y=-1$ rod. In this event, we recover the usual five-dimensional Schwarzschild black hole with $S^3$ horizon topology (see Appendix B).

To obtain a black lens with horizon topology $L(n,1)$, we set the second component of (\ref{orientation}) to be:
\be
\label{lens_condition1}
\frac{2ac}{1-c-a^2(1+c)}=n\,,
\ee
for positive integer $n$. This can be solved in terms of $a$ as 
\be
a=\frac{\pm\sqrt{c^2+n^2(1-c^2)}-c}{n(1+c)}\,.
\ee
It can be shown that the solution with positive sign satisfies $0<a<\sqrt{(1-c)/(1+c)}$; we call this Range I. On the other hand, the solution with negative sign satisfies $-1<a<-\sqrt{(1-c)/(1+c)}$; we call this Range II. It is instructive to plot the left-hand side of (\ref{lens_condition1}) against $a$ (for fixed $c$) to see these two ranges, as in Fig.~2. Note that in Range I, $n$ takes integer values in the interval $(0,\infty)$; while in Range II, $n$ takes integer values in the interval $(1,\infty)$.

\begin{figure}[t]
\begin{center}
\includegraphics{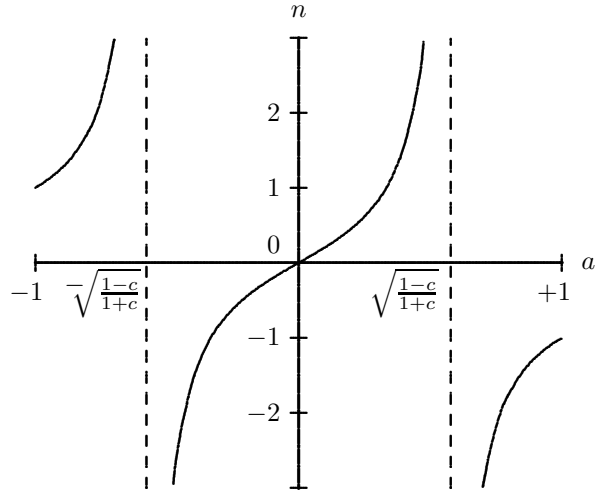}
\caption{Graph of the left-hand side of (\ref{lens_condition1}) against $a$, for fixed $c$}
\end{center}
\end{figure}

We now turn to a study of possible conical singularities in the space-time. 
Recall that the condition of asymptotic flatness requires $\psi$ and $\phi$ to have standard periodicity $2\pi$. This means that the coordinate associated with the Killing vector $\partial/\partial\tilde\psi\equiv n\partial/\partial\psi+\partial/\partial\phi$ that vanishes along the $x=1$ axis also has period $\Delta\tilde\psi=2\pi$. Now, the condition for the absence of a conical singularity along this axis is
\be
\label{Harmark_condition}
\Delta\tilde\psi=\frac{2\pi}{m}\,,
\ee
where $m$ is defined such that $2\pi/m$ is equal to the right-hand side of Eq.~(3.10) in \cite{Harmark:2004}. In other words, we require that $m=1$. In the present case, this condition becomes
\be
\label{conical_condition1}
m^2\equiv\frac{(1-a^2)^2(1-c^2)}{[1-c-a^2(1+c)]^2}=1\,.
\ee
Solving it gives
\be
\label{conical_condition1_solns}
a=\pm\sqrt[4]{(1-c)/(1+c)}\,.
\ee
The positive solution does not lie in Range I, so the conical singularity cannot be eliminated for any $a$ in this range; indeed, it can be shown that $m^2>n^2$. However, the negative solution of (\ref{conical_condition1_solns}) lies in Range II, so the conical singularity can be eliminated for this particular value of $a$. Imposing the two conditions (\ref{lens_condition1}) and (\ref{conical_condition1}) simultaneously, we see that it corresponds to the solution
\be
a=\frac{\sqrt{n^2-4}-n}{2}\,,\qquad 
c=\frac{1-a^4}{1+a^4}\,,
\ee
and is only applicable for $n\geq3$. (The $n=2$ case is excluded, as it implies $c=0$, which means that there is no longer a black lens present in the space-time.) The allowed range of $a$ for this solution is $-1<a<0$.

While it may seem from the preceding result that Range II is the more appropriate range to consider, all solutions in this range unfortunately suffer from the following pathology: It can be checked that the value of $H(x,y)$ is zero on a closed surface which separates the point $(x,y)=(1,-1)$ from the rest of the space-time, including the horizon and asymptotic infinity. Since the curvature invariant $R_{abcd}R^{abcd}\sim H(x,y)^{-6}$, this surface is a singular one; moreover, it is nakedly singular since it is not enclosed by the event horizon. Since this surface intersects the $y=-1$ and $x=1$ axes, it has an $L(1,n)=S^3$ spherical topology. On the other hand, it can be checked that for Range I, $H(x,y)$ is positive everywhere in the space-time outside the event horizon, so this naked singularity does not exist. For either range, if we extend the coordinate range below the horizon $y<-1/c$, we would also find a curvature singularity at $y\rightarrow-\infty$.

At this stage, let us explicitly examine the horizon geometry of our solution to confirm the above interpretation that it has an $L(n,1)$ lens-space topology. Our analysis will follow that of \cite{Lu:2008}. From (\ref{metric}), its metric on a constant time slice is given by
\ba
\label{metric2}
\dif s^2_\mathrm{H}&=&\frac{2\kappa^2}{(1-a^2)(1-c)(1+cx)}\Bigg\{c^2(1-c)^2\frac{\dif x^2}{G(x)}-2ac(1+x)(1+cx)[1-c-a^2(1+c)]\dif\psi\dif\phi\cr
&&\hskip1.8in+\left((1+c)[1-c-a^2(1+cx)]^2+a^2c^2(1-c)(1-x^2)\right)\dif\psi^2\cr
&&\hskip1.8in+c^2(1+x)[(1-c)(1-x)+a^2(1+c)(1+x)]\,\dif\phi^2\Bigg\}\,.
\ea
We now introduce new azimuthal coordinates $\tilde\phi$ and $\tilde\psi$, chosen such that the Killing vectors $\ell_1$ and $\ell_2$ that vanish at $x=-1$ and $x=1$, respectively, are simply given by
\be
\label{l1l2}
\ell_1=\frac{\partial}{\partial\tilde\phi}\,,\qquad\ell_2=\frac{\partial}{\partial\tilde\psi}\,.
\ee
These coordinates are related to $\phi$ and $\psi$ by
\be
\label{identifications}
\tilde\phi=\phi-\frac{1}{n}\,\psi\,,\qquad\tilde\psi=\frac{1}{n}\,\psi\,,
\ee
and can be seen to have period $2\pi$. If we recall that $n$ is given by the left-hand side of (\ref{lens_condition1}), the metric (\ref{metric2}) can be written in terms of these coordinates in either of the following two forms:
\ba
\label{metric3}
\dif s^2_\mathrm{H}&=&\frac{2\kappa^2c^2}{(1-a^2)(1+cx)}\Bigg[(1-c)\frac{\dif x^2}{G(x)}
+\frac{4a^2(1+x)}{g_1(x)}\,\dif\tilde\phi^2+(1-x)m^2g_1(x)\,(\dif\tilde\psi+f_1(x)\,\dif\tilde\phi)^2\Bigg]\,,\cr
\dif s^2_\mathrm{H}&=&\frac{2\kappa^2c^2}{(1-a^2)(1+cx)}\Bigg[(1-c)\frac{\dif x^2}{G(x)}
+\frac{4a^2m^2(1-x)}{g_2(x)}\,\dif\tilde\psi^2+(1+x)g_2(x)\,(\dif\tilde\phi+f_2(x)\,\dif\tilde\psi)^2\Bigg]\,.\cr
&&
\ea
Here, we have defined
\ba
g_1(x)&=&\frac{1-c}{1+c}\,(1+x)+a^2(1-x)\,,\qquad f_1(x)=\frac{(1-a^2)(1+x)}{m^2g_1(x)}\,,\cr
g_2(x)&=&(1-x)+a^2\frac{1+c}{1-c}\,(1+x)\,,\qquad f_2(x)=\frac{(1-a^2)(1-x)}{g_2(x)}\,,
\ea
and $m^2$ is given by the left-hand side of (\ref{conical_condition1}).
Both metrics in (\ref{metric3}) resemble that of a 3-sphere, albeit a squashed one, with degenerations occuring at $x=\pm1$. 
We can examine the vicinity of the `north pole' $x=-1$ by introducing the new coordinate $r_1$, such that $x=-1+r_1^2$. For small $r_1$, the first metric in (\ref{metric3}) reduces to
\be
\dif s^2_\mathrm{H}\rightarrow\frac{4\kappa^2c^2}{(1-a^2)(1-c)}(\dif r_1^2+r_1^2\,\dif\tilde\phi^2)+2\kappa^2(1-a^2)(1+c)n^2\,\dif\tilde\psi^2\,.
\ee
On the other hand, we can examine the vicinity of the `south pole' $x=1$ by introducing the new coordinate $r_2$, such that $x=1-r_2^2$. For small $r_2$, the second metric in (\ref{metric3}) reduces to
\be
\dif s^2_\mathrm{H}\rightarrow\frac{4\kappa^2c^2(1-c)}{(1-a^2)(1+c)^2}(\dif r_2^2+m^2r_2^2\,\dif\tilde\psi^2)+2\kappa^2(1-a^2)(1+c)\frac{n^2}{m^2}\,\dif\tilde\phi^2\,.
\ee
Note that there will be a conical singularity at the point $r_2=0$ unless $m=\pm1$, which is consistent with our interpretation above. 
Thus, the local behaviour of the horizon metric at these two poles (modulo the possible presence of the conical singularity) is similar to the standard metric on $S^3$. However, there are some global differences resulting from (\ref{identifications}). Since $\psi$ has period $2\pi$, it follows from (\ref{identifications}) that identifications should be made under the operation:
\be
\tilde\phi\rightarrow\tilde\phi-\frac{2\pi}{n}\,,\qquad\tilde\psi\rightarrow\tilde\psi+\frac{2\pi}{n}\,.
\ee
These are precisely the identifications of $S^3$ required to turn it into the lens space $L(n,1)$ \cite{Hollands:2007,Lu:2008}, thus confirming that the horizon has this topology.

It is instructive to similarly examine the geometry of the space near the point $(x,y)=(1,-1)$, where the $x=1$ and $y=-1$ axes meet up. We introduce azimuthal coordinates $\phi'$ and $\psi'$, chosen such that the Killing vectors $\ell_2$ and $\ell_3$ that vanish at $x=1$ and $y=-1$ are given by
\be
\ell_2=\frac{\partial}{\partial\phi'}\,,\qquad\ell_3=\frac{\partial}{\partial\psi'}\,.
\ee
These coordinates are related to $\phi$ and $\psi$ by
\be
\phi'=\phi\,,\qquad\psi'=\psi-n\phi\,,
\ee
and can be seen to have period $2\pi$. 
If we introduce the new coordinates $r$ and $\theta$ by
\be
x=1-(1+c)\,r^2\sin^2\theta\,,\qquad y=-1-(1-c)\,r^2\cos^2\theta\,,
\ee
then the spatial part of (\ref{metric}) in the region of small $r$ is
\be
\dif s^2=\frac{\kappa^2(1+c)[1-c-a^2(1+c)]}{1-a^2}\left[\dif r^2+r^2\left(\dif\theta^2+m^2\sin^2\theta\,\dif\phi'^2+\cos^2\theta\,\dif\psi'^2\right)\right].
\ee
This is just a flat-space geometry with a conical singularity along the $\theta=0$ ($x=1$) axis in general. The form of this geometry is exactly the same as that of the more familiar case of the black ring. In particular, there is no orbifold singularity at the origin $r=0$. For $a$ taking values in Range II however, this metric has the wrong signature, and is an indication of the fact that there are closed time-like curves (CTCs) in this region. 

Now, the requirement for the absence of CTCs is that the $2\times2$ metric $g_{ij}$, $i,j=\psi,\phi$, be positive semi-definite. This is equivalent to checking that its determinant and one of the diagonal components, say $g_{\psi\psi}$, are non-negative. These two quantities can be read off from (\ref{metric}), and we have checked that the following results hold: For $a$ taking values in Range I, there are no CTCs anywhere in the space-time outside the horizon. For $a$ taking values in Range II, there are no CTCs outside the horizon and naked singularity; however, CTCs do exist in the region inside the naked singularity, as we have seen above in the vicinity of $r=0$. Fortunately, they are not a concern for us as this region is not accessible to observers outside the naked singularity.

Finally, we note that the entropy (from area) and temperature (from surface gravity) of the black-lens horizon are given by the expressions:
\be
\label{entropy}
S=\frac{4\pi^2\kappa^3c^2}{G}\sqrt{\frac{2}{(1-a^2)(1+c)}}\,,\qquad
T=\frac{1}{8\pi\kappa c}\sqrt{2(1-a^2)(1+c)}\,.
\ee
It follows that the Smarr relation:
\be
\frac{2}{3}\,M=TS\,,
\ee
is satisfied by the static black lens. When $\kappa$ and $c$ are kept fixed (so that mass $M$ is fixed), observe from Fig.~2 that solutions in Range I have a value of $a$ that increases with $n$. From (\ref{entropy}), it follows that $S$ also increases with $n$, so black lenses with larger $n$ are entropically favoured. The configuration in this range with the highest entropy is the $n\rightarrow\infty$ limiting case of the Schwarzschild black hole. On the other hand, it can be seen that solutions in Range II have an entropy that decreases as $n$ is increased. In this case, black lenses with smaller $n$ are entropically favoured. Note also that solutions in Range II have higher entropy than solutions in Range I.

To summarise, the black lens solution (\ref{metric}) can be divided into two ranges I and II depending on the value of $a$, which exhibit rather different properties outside the horizon.  All solutions in Range I possess a conical singularity along the $x=1$ axis, but are otherwise regular and well-behaved. Included in this range are all positive values of $n$. For the case $n=1$ [corresponding to $a=(1-c)/(1+c)$], we recover a black hole with $L(1,1)=S^3$ horizon topology. It differs from the usual Schwarzschild black hole because of the presence of the conical singularity in the space-time; this solution will be revisited in Appendix B. For the case $n=2$, we have a black lens with $L(2,1)=\mathbb{R}P^3$ event-horizon topology.

Solutions in Range II also in general possess a conical singularity along the $x=1$ axis, although it can be eliminated for a particular value of $a$ in this range and with $n\geq3$. However, all solutions in this range possess a naked singularity with spherical topology surrounding the point $(x,y)=(1,-1)$. A similar situation was found in the static black lens solution of \cite{Evslin:2008}, which actually contains {\it two\/} such singularities. It was conjectured in \cite{Evslin:2008} that adding angular momentum might eliminate such singularities, and we shall revisit this issue in the following section when we add a single asymptotic angular momentum to our black lens solution.

\newsection{Rotating black lens}

The derivation of the single-rotating black lens solution begins with the derivation of the single-rotating black ring solution using the inverse scattering method. This was originally done in \cite{Iguchi:2006,Tomizawa:2006} using a two-soliton transformation, although we used a simpler version of it involving just one soliton similar to the one described in \cite{Elvang:2007,Emparan:2008}.\footnote{The main difference is that we placed the negative density rod to the right of the horizon, rather than to the left as in Fig.~9(a) of \cite{Emparan:2008}. It is also possible to obtain the same solution with the latter choice, although the solitons to be subsequently removed and added back would have to be at different locations and have different BZ vectors.} To then make the finite space-like rod rotate to the correct orientation, we need to perform the following operations: Borrowing from the labelling of Fig.~1, first remove solitons from $z_1$ and $z_3$, both with BZ vector $(0,0,1)$. Then add back the two solitons, with BZ vectors $(C_2,0,1)$ and $(0,C_1,1)$, respectively. Here, $C_1$ and $C_2$ are appropriately chosen constants; in particular, $C_1$ turns out to be proportional to the parameter $a$ introduced above in the static case. A final change in coordinates is needed to bring the solution into an asymptotically flat form. The constant $C_2$ is chosen to ensure that the finite space-like rod has no time-like component in this final solution.\footnote{This is equivalent to demanding that there are no Dirac--Misner string singularities along this axis.}

In C-metric type coordinates, the metric we obtain is 
\ba
\label{metric1}
\dif s^2&=&-\frac{H(y,x)}{H(x,y)}(\dif t-\omega_\psi\,\dif\psi-\omega_\phi\,\dif\phi)^2
-\frac{F(x,y)}{H(y,x)}\,\dif\psi^2+2\frac{J(x,y)}{H(y,x)}\,\dif\psi\dif\phi\cr
&&+\frac{F(y,x)}{H(y,x)}\,\dif\phi^2+\frac{\kappa^2H(x,y)}{2(1-a^2)(1-b)^3(x-y)^2}\left(\frac{\dif x^2}{G(x)}-\frac{\dif y^2}{G(y)}\right),
\ea
where
\ba
\omega_\psi&=&\frac{2\kappa}{H(y,x)}\sqrt{\frac{2b(1+b)(b-c)}{(1-a^2)(1-b)}}\,
(1-c)(1+y)\{2[1-b-a^2(1+bx)]^2(1-c)\cr
&&\hskip2.1in~-a^2(1-a^2)b(1-b)(1-x)(1+cx)(1+y)\}\,,\cr
\omega_\phi&=&\frac{2\kappa}{H(y,x)}\sqrt{\frac{2b(1+b)(b-c)}{(1-a^2)(1-b)}}\,
a(1-c)(1+x)^2(1+y)[a^4(1+b)(b-c)\cr
&&\hskip2.1in~+a^2(1-b)(-b+cb+2c)-(1-b)^2c]\,.
\ea 
The functions $G$, $H$, $F$ and $J$ are defined as
\ba
G(x)&=&(1-x^2)(1+cx)\,,\cr
H(x,y)&=&4(1-b)(1-c)(1+bx)\{(1-b)(1-c)-a^2[(1+bx)(1+cy)+(b-c)(1+y)]\}\cr
&&+a^2(b-c)(1+x)(1+y)\{(1+b)(1+y)[(1-a^2)(1-b)c(1+x)+2a^2b(1-c)]\cr
&&-2b(1-b)(1-c)(1-x)\}\,,\cr
F(x,y)&=&\frac{2\kappa^2}{(1-a^2)(x-y)^2}\,\Big[4(1-c)^2(1+bx)[1-b-a^2(1+bx)]^2G(y)\cr
&&~-a^2G(x)(1+y)^2\Big([1-b-a^2(1+b)]^2(1-c)^2(1+by)-(1-a^2)(1-b^2)\times\cr
&&~\times(1+cy)\{(1-a^2)(b-c)(1+y)+[1-3b-a^2(1+b)](1-c)\}\Big)\Big]\,,\cr
J(x,y)&=&\frac{4\kappa^2a(1-c)(1+x)(1+y)}{(1-a^2)(x-y)}\,[1-b-a^2(1+b)][(1-b)c+a^2(b-c)]\times\cr
&&~\times[(1+bx)(1+cy)+(1+cx)(1+by)+(b-c)(1-xy)]\,.
\ea
The coordinates take the same ranges as in the previous section, while the parameters satisfy $0<c\leq b<1$ and $-1<a<1$. Note that the metric is invariant under the action $a\rightarrow-a$ and $\phi\rightarrow-\phi$.
The ADM mass and angular momenta of this space-time can be calculated to be
\be
\label{mass}
M=\frac{3\pi\kappa^2b(1-c)}{2G(1-b)}\,,\qquad J_\psi=\frac{\pi\kappa^3\sqrt{2(1-a^2)b(1+b)(b-c)}(1-c)}{G(1-b)^{3/2}}\,,\qquad J_\phi=0\,.
\ee
As desired, we have a space-time with angular momentum in a single direction, namely the $\psi$ direction. The static limit is recovered when $b=c$, and (\ref{metric1}) reduces to the previous solution (\ref{metric}). The background limit is recovered when $b,c\rightarrow0$, and this will be examined in more detail in Appendix A.

The orientation of the $y=-1$ rod is $(0,1,0)$, while that of the $x=-1$ rod is $(0,0,1)$, as expected. The orientation of the finite space-like rod at $x=1$ is
\be
\left(0,\frac{2a[(1-b)c+a^2(b-c)]}{[1-b-a^2(1+b)](1-c)},1\right).
\ee
Two special cases can immediately be read off from this result. One is when $a=0$, in which the $x=1$ rod is parallel to the $x=-1$ rod. In this event, we recover the Emparan--Reall black ring with parameters $b$ and $c$ as introduced in \cite{Harmark:2004}. The other is when $a=\pm\sqrt{(1-b)/(1+b)}$; in which the $x=1$ rod is joined up to the $y=-1$ rod in the same direction. In this event, we recover the single-rotating Myers--Perry black hole (see Appendix B).

To obtain a black lens with horizon topology $L(n,1)$, we set
\be
\label{lens_condition2}
\frac{2a[(1-b)c+a^2(b-c)]}{[1-b-a^2(1+b)](1-c)}=n\,,
\ee
for positive integer $n$. To study the solutions of this equation, it is again instructive to plot the left-hand side of (\ref{lens_condition2}) against $a$ (for fixed $b$ and $c$). The graph obtained is qualitatively similar to that in Fig.~2, except that now the vertical asymptotes are located at $a=\pm\sqrt{(1-b)/(1+b)}$. There are again two ranges of solutions to consider: the first $0<a<\sqrt{(1-b)/(1+b)}$, which we call Range I; and the second $-1<a<-\sqrt{(1-b)/(1+b)}$, which we call Range II. Note that in Range I, $n$ takes integer values in the interval $(0,\infty)$; while in Range II, $n$ takes integer values in the interval $(1,\infty)$.

The finite time-like rod at $y=-1/c$ represents the event horizon of the black lens. Its orientation can be calculated to be
\be
\label{angular_velocity}
\left(1,\frac{1}{\kappa}\sqrt{\frac{(1-b)(b-c)}{2(1-a^2)b(1+b)}}\frac{1}{1-c},\frac{1}{2\kappa}\sqrt{\frac{(1-b)(b-c)}{2(1-a^2)b(1+b)}}\frac{a[1-b-a^2(1+b)]}{(1-b)c+a^2(b-c)}\right).
\ee
The second component represents the angular velocity of the horizon in the $\psi$ direction, while the third component represents its angular velocity in the $\phi$ direction. The ratio of these two quantities is, in fact, $n/a^2$. Thus, the event horizon of the black lens is rotating in both the $\psi$ and $\phi$ directions, although not independently. This is in contrast to the situation at asymptotic infinity, in which only the angular momentum in the $\psi$-direction survives.

Let us now turn to a study of possible conical singularities in the space-time.
With $\psi$ and $\phi$ taking the standard periodicity $2\pi$, the coordinate $\tilde\psi$ associated with the $x=1$ axis also has period $\Delta\tilde\psi=2\pi$. Since the condition for the absence of a conical singularity along this axis is given by (\ref{Harmark_condition}), we require $m=1$. In the present case, this condition becomes
\be
\label{conical_condition2}
m^2\equiv\frac{(1-a^2)^2(1-b)^3(1+c)^2}{[1-b-a^2(1+b)]^2(1+b)(1-c)^2}=1\,.
\ee
As in the static case, it is possible to show that this condition cannot be satisfied for any $a$ in Range I. Indeed, solving (\ref{lens_condition2}) in terms of $b$ or $c$ and substituting it into the left-hand side of (\ref{conical_condition2}) shows that $m^2>(n+a)^2$.

Thus, the only possibility for the conical singularity to be eliminated lies in Range II. It turns out to be simpler to solve the conditions (\ref{lens_condition2}) and (\ref{conical_condition2}) simultaneously in terms of $(b,c)$, rather than $(a,b)$ or $(a,c)$. We obtain the solution:
\be
\label{bc}
b = \frac{n(n+2a)}{n^2+2na+2a^2}\,,\qquad
c = \frac{n(1-n^2-3na-3a^2)}{(n+2a)(1-n^2-na-a^2)}\,.
\ee
The requirement that $b\geq c$ then implies that $a$ takes values in the range
\be
\label{bc_range}
\frac{\sqrt{n^2-4}-n}{2}\leq a<0\,,
\ee
with the lower bound corresponding to the static case $b=c$. Note that unlike the static case, the $n=2$ solution is a valid one when there is rotation. The static limit of this particular solution forces $b,c\rightarrow0$, and the black lens disappears leaving just the background space-time (see Appendix A). 

Unfortunately, all solutions in Range II suffer from the same pathology as in the static case; namely, the value of $H(x,y)$ vanishes on a closed surface of spherical topology that surrounds the point $(x,y)=(1,-1)$, separating it from the rest of the space-time, including the horizon and asymptotic infinity. On this surface, the curvature invariant $R_{abcd}R^{abcd}$ diverges, so it is a nakedly singular one. This singularity does not exist in Range I.
For either range, if we extend the coordinate range below the horizon $y<-1/c$, there is also a curvature singularity at $y\rightarrow-\infty$, $x\rightarrow-1$.

When rotation is present, there will be an ergoregion in the space-time where the Killing vector $\partial/\partial t$ changes from being time-like to space-like. It is bounded by the closed surface on which the value of $H(y,x)$ vanishes. For solutions in Range I, it is possible to show that this surface completely encloses the event horizon, and only intersects the $x=-1$ and $x=1$ axes. Thus, this surface has the same $L(n,1)$ topology as the event horizon. For solutions in Range II with sufficiently small values of $a^2$, there will continue to be an ergoregion with surface topology $L(n,1)$ enclosing the event horizon. However, another separate ergoregion appears enclosing the naked singularity, with an $S^3$ surface topology since it intersects the $y=-1$ and $x=1$ axes. For larger values of $a^2$ (which includes the case when the conical singularity is eliminated), the two ergoregions in fact merge into a single one that encloses {\it both\/} the event horizon and the naked singularity, as well as the finite axis $x=1$. Its surface intersects the $x=-1$ and $y=-1$ axes, and so it has an $S^3$ topology.

Now, it is possible to analyse the horizon geometry as was done for the static case, to verify that it has an $L(n,1)$ lens-space topology. The details are similar; in particular, the required transformation has the same form as (\ref{identifications}). We will not repeat the analysis here. Similarly, it can be shown that the geometry near the point $(x,y)=(1,-1)$ is just flat space with a conical singularity along the $x=1$ axis. However, when $a$ takes values in Range II, there appears to be CTCs in this region.

We can check the possible existence of CTCs in the space-time (\ref{metric1}), in the same manner as the static case. Since the expressions are a lot more complicated in this case, we have resorted to numerical analysis to do so. When $a$ takes values in Range I, no CTCs were found outside the horizon despite extensive checks. However, when $a$ takes values in Range II, CTCs were found not only inside the naked singularity, but also {\it outside\/} it when there is rotation present. These CTCs tend to occur very close to the surface of the naked singularity, and so appear to be more a feature of the naked singularity rather than the black lens itself. 

Finally, we note the following expressions for the entropy and temperature of the black-lens horizon:
\ba
S&=&\frac{4\pi^2\kappa^3}{G}\sqrt{\frac{2bc(1+b)}{1-a^2}}\frac{(1-c)[(1-b)c+a^2(b-c)]}{(1-b)^2(1+c)}\,,\cr
T&=&\frac{1}{8\pi\kappa}\sqrt{\frac{2c(1-a^2)}{b(1+b)}}\frac{(1-b)^2(1+c)}{(1-c)[(1-b)c+a^2(b-c)]}\,.
\ea
It can then be checked that the rotating black lens satisfies the Smarr relation:
\be
\frac{2}{3}\,M=TS+\Omega_\psi J_\psi+\Omega_\phi J_\phi\,,
\ee
where the angular velocities of the horizon, $\Omega_\psi$ and $\Omega_\phi$, are given by the second and third components of (\ref{angular_velocity}), respectively. 
We also note that the Komar mass and angular momenta evaluated at the horizon of the black lens agree with the asymptotic quantities in (\ref{mass}). This implies that the conical singularity and/or naked singularity do not contribute to the total ADM mass and angular momentum of the space-time.

To summarise, the black lens solution (\ref{metric1}) can be divided into two ranges I and II, as defined above, which exhibit different properties.  All solutions in Range I possess a conical singularity along the $x=1$ axis, but are otherwise regular and well-behaved. Included in this range are all positive values of $n$. For the case $n=1$, we recover a rotating black hole with $L(1,1)=S^3$ horizon topology, with a conical singularity attached to it. This solution will be revisited in Appendix B. For the case $n=2$, we have a rotating black lens with $L(2,1)=\mathbb{R}P^3$ horizon topology.

Solutions in Range II also in general possess a conical singularity along the $x=1$ axis, although it can be eliminated for a particular value of $a$ in this range and with $n\geq2$. However, all solutions in this range possess a naked singularity with spherical topology surrounding the point $(x,y)=(1,-1)$. Thus, the introduction of a single rotation to the black lens does not remove this singularity, as was hoped for in \cite{Evslin:2008}. Moreover, the rotation causes CTCs to appear just outside the surface of the naked singularity. If one does not desire the presence of CTCs with its associated paradoxes, then it would appear that Range I solutions are the more appropriate ones to consider.

\newsection{Discussion}

The main results we have obtained are as follows: We have derived the metric for an asymptotically flat black lens with $L(n,1)$ event-horizon topology, with asymptotic angular momentum in one direction. Unfortunately, we have found that this space-time cannot be made completely regular. One either has to have a conical singularity attached to the black-lens event horizon, or a spherical naked singularity away from the event horizon. The latter interpretation was adopted by Evslin \cite{Evslin:2008}, who argued that since this singularity is isolated from the event horizon, it may somehow be eliminated locally without affecting the black lens itself. One of the results we have found is that introducing a single angular momentum does not seem able to eliminate it.

An obvious extension of this study would be to include angular momentum in the other azimuthal direction. We have in fact used the inverse scattering method to construct a black lens solution with two independent angular momenta. It contains, as special cases, both the double-rotating black ring \cite{Pomeransky:2006} as well as the single-rotating black lens (\ref{metric1}). Unfortunately, this solution has a very complicated form which we will not present here. However, we have analysed its properties numerically, and it does not seem to be possible to eliminate the conical and naked singularities simultaneously, while at the same time maintaining a positive ADM mass for the black lens. We hope to present these results in more detail elsewhere.

It may well turn out that completely regular black lenses do not exist, and that either conical or naked singularities are unavoidable. Of these two possibilities, we actually prefer the scenario containing conical singularities. The presence of a conical singularity (which in this case can be seen to be a conical {\it excess\/}, corresponding to what is also known as a strut singularity) is physically needed to balance the gravitational self-attraction of the black lens, something that the centrifugal force from the rotation seems unable to do alone. Such conical singularities are also quite common in other black hole solutions in general relativity, especially in space-times containing multiple black holes, all of which are considered legitimate space-times.

There are other reasons to prefer the scenario with conical singularities rather than naked singularities. As we have found, the introduction of angular momentum to the space-time causes closed time-like curves to appear near the naked singularity, but otherwise seem to be absent in those solutions without naked singularities. The class of black lens solutions containing just conical singularities (Range I as defined above) also has the feature that the familiar black ring solution emerges as a limiting case, and so can be regarded as the natural generalisation of the black ring. It follows that this class of solutions possesses some appealing properties in common with the black ring solution, such as the existence of a well-defined black hole limit (see Appendix B).

What about black lenses with more general event-horizon topology $L(p,q)$, where $p,q$ are coprime integers? Actually, this solution is still given by (\ref{metric1}), if we set $n=p/q$ in (\ref{lens_condition2}). The condition for there to be no conical singularities along $x=1$ is then $m=1/q$, instead of $m=1$ as in (\ref{conical_condition2}). It turns out that the analysis in Sec.~3 is still valid in this case, and that there will be either a conical singularity or a naked singularity [with topology $L(q,p)$] in the space-time. Furthermore, there will be a $\mathbb{Z}_q$ orbifold singularity at the point $(x,y)=(1,-1)$ when $q>1$. This orbifold singularity may be resolved by introducing a second black lens with horizon topology $L(q,p)$ at this point.

In a recent paper \cite{Lu:2008}, Lu et al.\ actually rediscovered and provided an alternative interpretation of the metric (\ref{metric}) which has {\it no\/} conical or naked singularities. They were able to satisfy both conditions (\ref{lens_condition1}) and (\ref{Harmark_condition}) simultaneously, by giving up the condition of asymptotic flatness: that $\psi$ and $\phi$ have period $2\pi$. What they found was a static black lens solution with horizon topology $L(n,m)$, asymptotic to a locally flat space-time whose spatial sections are in fact lens spaces $L(m,n)$. We expect that the metric (\ref{metric1}) can similarly be interpreted as a rotating $L(n,m)$ black lens with an $L(m,n)$ asymptotic structure, for appropriate reidentifications of $\psi$ and $\phi$. 

It is also possible to consider charged versions of our rotating black lens solution, for example in the context of five-dimensional minimal supergravity. Such a solution can be constructed using standard solution-generating techniques (see, e.g., \cite{Lu:2008b}), and would be expected to carry both an electric charge and a (non-conserved) magnetic dipole charge. It would be interesting to construct and examine the properties of this solution. However, we note, somewhat disappointedly, that asymptotically flat {\it supersymmetric\/} black lenses have been proved not to exist in five-dimensional minimal supergravity \cite{Reall:2002}.

At this point, one may be tempted to ponder about asymptotically flat black holes with non-spherical horizon topologies in six or higher dimensions. Unfortunately, most of the methods relied upon in this paper: the generalised Weyl formalism \cite{Emparan:2001a} and the concept of rod structures \cite{Harmark:2004}, the inverse scattering method \cite{Belinsky:1971,Belinsky:1979,Belinski:2001}, etc., will no longer be applicable. This is due to the simple fact that black holes in six or higher dimensions do not have the requisite number of commuting Killing vectors, and complete integrability of the Einstein equations is no longer assured.  Thus, finding higher-dimensional analogues of black rings or black lenses would probably require a radically different approach. This is certainly a worthwhile and challenging problem to be left for the future.

\appendix

\newsection{Background space-time}

The background space-time for the static black lens (\ref{metric}), or more generally the rotating black lens (\ref{metric1}), can be uniquely determined after making the following reasonable assumptions: Firstly, we should take $c\rightarrow0$ to eliminate the horizon; in the context of Fig.~1, this corresponds to taking the limit $z_1\rightarrow z_2$. Secondly, the mass and angular momentum of the background space-time should vanish. From the relevant expressions in (\ref{mass}), it follows that we should also take $b\rightarrow0$ while keeping $b\geq c$. On the other hand, the non-negative integer $n$ should be fixed to maintain the orientation of the finite space-like rod. For solutions in Range I, it follows from (\ref{lens_condition2}) that $a$ takes the form 
\be
a=1-\frac{1+n}{n}\,b+O(b^2)\,,
\ee
in this limit. The black lens metric (\ref{metric1}) then reduces to
\ba
\label{background}
\dif s^2=-\dif t^2
&+&\frac{\kappa^2}{(1+n)(x-y)^2}\Bigg\{
[2-n(x+y)]\left(\frac{\dif x^2}{1-x^2}-\frac{\dif y^2}{1-y^2}\right)\cr
&&+\frac{1-x^2}{2-n(x+y)}\left[(2+n(1-y))\dif\phi-n(1+y)\dif\psi\right]^2\cr
&&-\frac{1-y^2}{2-n(x+y)}\left[(2+n(1-x))\dif\psi-n(1+x)\dif\phi\right]^2\Bigg\}\,,
\ea
with $-\infty<t<\infty$, $-1\leq x\leq 1$, $-\infty<y\leq-1$, and $\psi,\phi$ having periodicity $2\pi$ to ensure asymptotic flatness. Note that this background depends only on the parameters $\kappa$ and $n$, as expected. It has a non-vanishing curvature if $n\neq0$. 

It can be verified that this space-time contains three axes: the two usual semi-infinite axes at $x=-1$ and $y=-1$, and a finite one at $x=1$ with orientation $(0,n,1)$. There is in general a conical singularity along the latter axis, since a calculation of $m^2$ in (\ref{conical_condition2}) shows that it has value $(1+n)^2$ in the background limit. On the other hand, we have checked that there are no CTCs present in this space-time. 

There are clearly two special points in the space-time that deserve attention. The first is where the $x=1$ and $y=-1$ axes meet up. The region around this point was already examined in Sec.~2, and the details are largely similar in this case. The second is where the $x=-1$ and $x=1$ axes meet up. Let us now examine the region around this point. 
We change to new coordinates $(r,\theta)$ as follows:
\be
x=\cos2\theta\,,\qquad y=-\frac{2}{r^2}\,,
\ee
where $0\leq\theta\leq\pi/2$. The point in question is then located at $r=0$. For small $r$, the spatial part of the metric (\ref{background}) becomes
\be
\dif s^2=\frac{2\kappa^2n}{1+n}\Bigg\{\dif r^2+r^2\Bigg[\dif\theta^2+\frac{(1+n)^2}{n^2}\sin^2\theta\,\dif\psi^2+\cos^2\theta\,\bigg(\dif\phi-\frac{1}{n}\,\dif\psi\bigg)^2\Bigg]\Bigg\}\,,
\ee
which is just a flat-space geometry. We may introduce azimuthal coordinates $\tilde\phi$ and $\tilde\psi$, such that the Killing vectors $\ell_1$ and $\ell_2$ that vanish at $\theta=\pi/2$ and $\theta=0$ are given by (\ref{l1l2}). They are related to $\phi$ and $\psi$ by
\be
\tilde\phi=\phi-\frac{1}{n}\,\psi\,,\qquad\tilde\psi=\frac{1}{n}\,\psi\,.
\ee
When $n\geq2$, this transformation does not have unit determinant, and it follows that there is a $\mathbb{Z}_n$ orbifold singularity at $r=0$. There is no orbifold singularity when $n=1$, which can also be seen from the fact that in this case, the two points where the axes meet up are mirror images of each other. In general, there is also a conical singularity [with excess angle $2\pi n/(1+n)$] along the $\theta=0$ axis.

On the other hand, for solutions in Range II, it follows from (\ref{lens_condition2}) that $a$ should take the form 
\be
a=-1+\frac{n-1}{n}\,b+O(b^2)\,,
\ee
in the background limit. In this case, the black lens metric (\ref{metric1}) then reduces to
\ba
\dif s^2=-\dif t^2
&+&\frac{\kappa^2}{(1-n)(x-y)^2}\Bigg\{
[2+n(x+y)]\left(\frac{\dif x^2}{1-x^2}-\frac{\dif y^2}{1-y^2}\right)\cr
&&+\frac{1-x^2}{2+n(x+y)}\left[(2-n(1-y))\dif\phi-n(1+y)\dif\psi\right]^2\cr
&&-\frac{1-y^2}{2+n(x+y)}\left[(2-n(1-x))\dif\psi-n(1+x)\dif\phi\right]^2\Bigg\}\,.
\ea
This background is actually related to the previous one (\ref{background}) under the transformation $n\rightarrow-n$ and either $\psi\rightarrow-\psi$ or $\phi\rightarrow-\phi$. It has the same rod structure as (\ref{background}); in particular, there is a $\mathbb{Z}_n$ orbifold singularity at the point where the $x=-1$ and $x=1$ axes meet up. There is in general a conical singularity along the $x=1$ axis, as well as a naked singularity with spherical topology located at points where $2+n(x+y)=0$. It can be checked that there are no CTCs outside this naked singularity. 

Now, if we restrict ourselves to the particular solution (\ref{bc}) in Range II which does not contain conical singularities, then we would require (\ref{conical_condition2}) to hold even in the background limit. A calculation shows that $m^2=(n-1)^2$, which means that this background can only be obtained for the special case of the $n=2$ solution.\footnote{Another way to see this is to plot $c$, as given in (\ref{bc}), against $a$ in the range (\ref{bc_range}). For $n>2$, the graphs do not touch the $c=0$ axis at any point.} As mentioned in Sec.~3, this corresponds to taking the static limit of this solution. However, there is still the spherical naked singularity in this background.

\newsection{Black-hole limits}

There are three limits in which black holes with spherical event horizon topology can be obtained from our black lens solution. In the interest of generality, we will only consider limits of the rotating black lens solution (\ref{metric1}) here. The black-hole limits of the static black lens can readily be obtained as special cases.

First, consider the case when $a=\pm\sqrt{(1-b)/(1+b)}$, in which the $x=1$ rod is joined up to the $y=-1$ rod in the same direction. To show that the metric (\ref{metric1}) is equivalent to the Myers--Perry black hole, we need to first transform to Weyl coordinates, and then to prolate spheroidal coordinates. The relevant formulae can be found in the appendices of \cite{Harmark:2004}. The coordinate transformation relating the C-metric coordinates $(x,y)$ to the prolate spheroidal coordinates $(\tilde x,\tilde y)$ is 
\ba
\label{transf1}
x&=&\frac{(1-c)R_1-(1+c)R_2-2R_3+2(1-c^2)\kappa^2}{(1-c)R_1+(1+c)R_2+2cR_3}\,,\cr
y&=&\frac{(1-c)R_1-(1+c)R_2-2R_3-2(1-c^2)\kappa^2}{(1-c)R_1+(1+c)R_2+2cR_3}\,,
\ea
where
\be
R_1=c\kappa^2(\tilde x+\tilde y)\,,\qquad 
R_2=c\kappa^2(\tilde x-\tilde y)\,,\qquad
R_3=\kappa^2\sqrt{c^2(\tilde x^2-1)(1-\tilde y^2)+(c\tilde x\tilde y-1)^2}\,.
\ee
Finally, we transform to Boyer--Lindquist coordinates $(r,\theta)$:
\be
\tilde x=\frac{2r^2}{\mu-\alpha^2}-1\,,\qquad\tilde y=\cos2\theta\,,
\ee
where $\mu$ and $\alpha$ are new parameters, related to $b$ and $c$ by
\be
\label{transf4}
b=\frac{\mu}{4\kappa^2+\alpha^2}\,,\qquad c=\frac{\mu-\alpha^2}{4\kappa^2}\,.
\ee
Under these transformations, (\ref{metric1}) becomes
\ba
\label{MP}
\dif s^2&=&-\frac{r^2-\mu+\alpha^2\cos^2\theta}{r^2+\alpha^2\cos^2\theta}\left(\dif t+\frac{\mu\alpha\sin^2\theta}{r^2-\mu+\alpha^2\cos^2\theta}\,\dif\psi\right)^2+r^2\cos^2\theta\,\dif\phi^2\cr
&&+(r^2+\alpha^2\cos^2\theta)\left(\frac{\dif r^2}{r^2-\mu+\alpha^2}+\dif\theta^2+\frac{r^2-\mu+\alpha^2}{r^2-\mu+\alpha^2\cos^2\theta}\,\sin^2\theta\,\dif\psi^2\right),
\ea
which is the familiar form of the Myers--Perry black hole \cite{Myers:1986} rotating in the $\psi$ direction. 

The second limit we shall consider is when the finite axis in the black lens space-time is shrunk to zero length --- $z_2\rightarrow z_3$ in the context of Fig.~1 --- while preserving its orientation. It turns out that we recover the Myers--Perry black hole only when $a$ takes values in Range I, which we recall contains the black ring as a limiting case. Indeed, the transformation we seek is similar to the one used in the black ring case \cite{Emparan:2004}: We define the parameters $\mu$ and $\alpha$ by
\be
\mu=\frac{4\kappa^2}{1-c}\,,\qquad \alpha^2=4\kappa^2\frac{b-c}{(1-c)^2}\,,
\ee
such that they remain finite in the limit $b,c\rightarrow1$ and $\kappa\rightarrow0$. In this limit, the relevant root of (\ref{lens_condition2}) has the form
\be
a=\frac{n}{2}(1-c)+O(1-c)^2.
\ee
If we transform to new coordinates $r$ and $\theta$ via the relations
\ba
x&=&-1+\left(1-\frac{\alpha^2}{\mu}\right)\frac{4\kappa^2\cos^2\theta}{r^2-(\mu-\alpha^2)\cos^2\theta}\,,\cr
y&=&-1-\left(1-\frac{\alpha^2}{\mu}\right)\frac{4\kappa^2\sin^2\theta}{r^2-(\mu-\alpha^2)\cos^2\theta}\,,
\ea
and rescale $t$:
\be
t\rightarrow\sqrt{\frac{4\kappa^2}{\mu-\alpha^2}}\,\,t\,,
\ee
then it can be checked that the metric (\ref{metric1}) indeed reduces to the Myers--Perry metric (\ref{MP}), up to an overall constant factor.

The third limit in which a black hole can be obtained from the black lens solution is when $n=1$. In this case, the topology of the horizon is also an $S^3$. However, it differs from the usual Myers--Perry black hole in that there is now a conical singularity attached to the black hole along the $x=1$ axis. One way to see this is to push the $y=-1$ axis to infinity by making the $x=1$ axis infinitely long, while preserving the latter's orientation. In the context of Fig.~1, this corresponds to taking $z_3\rightarrow\infty$. We first perform the transformations (\ref{transf1}) to (\ref{transf4}), and then take the limits $b,c\rightarrow0$ and $\kappa\rightarrow\infty$ such that $b\kappa^2$ and $c\kappa^2$ remain finite. In this limit, the relevant root of (\ref{lens_condition2}) has the form
\be
a=1-\frac{(1+n)\mu}{4n\kappa^2}+O\left(\frac{1}{\kappa^4}\right),
\ee
where we keep $n$ general for the time being.
If we rescale $t\rightarrow\sqrt{n/(1+n)}\,t$, then (\ref{metric1}) becomes, up to an overall constant factor,
\ba
\label{MP2}
\dif s^2&=&-\frac{r^2-\mu+\alpha^2\cos^2\theta}{r^2+\alpha^2\cos^2\theta}\left(\dif t+\frac{\mu\alpha\sin^2\theta}{r^2-\mu+\alpha^2\cos^2\theta}\frac{1+n}{n}\,\dif\psi\right)^2+r^2\cos^2\theta\,\left(\dif\phi-\frac{1}{n}\,\dif\psi\right)^2\cr
&&+(r^2+\alpha^2\cos^2\theta)\left(\frac{\dif r^2}{r^2-\mu+\alpha^2}+\dif\theta^2+\frac{r^2-\mu+\alpha^2}{r^2-\mu+\alpha^2\cos^2\theta}\frac{(1+n)^2}{n^2}\,\sin^2\theta\,\dif\psi^2\right).\cr&&
\ea
Now, we may introduce azimuthal coordinates $\tilde\phi$ and $\tilde\psi$, such that the Killing vectors $\ell_1$ and $\ell_2$ that vanish at $\theta=\pi/2$ and $\theta=0$ are given by (\ref{l1l2}). They are related to $\phi$ and $\psi$ by
\be
\tilde\phi=\phi-\frac{1}{n}\,\psi\,,\qquad\tilde\psi=\frac{1}{n}\,\psi\,.
\ee
In these coordinates, the space-time described by (\ref{MP2}) for the case $n=1$ can be seen to be just the Myers--Perry black hole rotating in the $\tilde\psi$ direction, but with a conical singularity (with excess angle $\pi$) along the $\theta=0$ axis. For $n\geq2$ however, this space-time is quotiented by $\mathbb{Z}_n$, so that (\ref{MP2}) describes a black lens with horizon topology $L(n,1)$ rotating in the $\tilde\psi$ direction, asymptotic to a locally flat space-time whose spatial sections are also lens spaces $L(n,1)$. Note that there is still a conical singularity [with excess angle $2\pi n/(1+n)$] along the $\theta=0$ axis.

\bigskip\bigskip

{\renewcommand{\Large}{\normalsize}
}

\begin{thebibliography}{99}


\bibitem{Emparan:2001}
  R.~Emparan and H.~S.~Reall,
  ``A rotating black ring solution in five dimensions,''
  Phys.\ Rev.\ Lett.\  {\bf 88} (2002) 101101
  [arXiv:hep-th/0110260].

\bibitem{Emparan:2006}
  R.~Emparan and H.~S.~Reall,
  ``Black rings,''
  Class.\ Quant.\ Grav.\  {\bf 23} (2006) R169
  [arXiv:hep-th/0608012].

\bibitem{Myers:1986}
  R.~C.~Myers and M.~J.~Perry,
  ``Black holes in higher dimensional space-times,''
  Annals Phys.\  {\bf 172} (1986) 304.

\bibitem{Cai:2001}
  M.~l.~Cai and G.~J.~Galloway,
  ``On the topology and area of higher dimensional black holes,''
  Class.\ Quant.\ Grav.\  {\bf 18} (2001) 2707
  [arXiv:hep-th/0102149].

\bibitem{Helfgott:2005}
  C.~Helfgott, Y.~Oz and Y.~Yanay,
  ``On the topology of black hole event horizons in higher dimensions,''
  JHEP {\bf 0602} (2006) 025
  [arXiv:hep-th/0509013].

\bibitem{Galloway:2005}
  G.~J.~Galloway and R.~Schoen,
  ``A generalization of Hawking's black hole topology theorem to higher
  dimensions,''
  Commun.\ Math.\ Phys.\  {\bf 266} (2006) 571
  [arXiv:gr-qc/0509107].

\bibitem{Hollands:2007}
  S.~Hollands and S.~Yazadjiev,
  ``Uniqueness theorem for 5-dimensional black holes with two axial Killing
  fields,''
  Commum.\ Math.\ Phys. (to be published) 
  [arXiv:0707.2775 [gr-qc]].

\bibitem{Hollands:2007b}
  S.~Hollands and S.~Yazadjiev,
  ``A uniqueness theorem for 5-dimensional Einstein--Maxwell black holes,''
  Class.\ Quant.\ Grav.\  {\bf 25} (2008) 095010 
  [arXiv:0711.1722 [gr-qc]].

\bibitem{Emparan:2001a}
  R.~Emparan and H.~S.~Reall,
  ``Generalized Weyl solutions,''
  Phys.\ Rev.\  D {\bf 65} (2002) 084025
  [arXiv:hep-th/0110258].

\bibitem{Harmark:2004}
  T.~Harmark,
  ``Stationary and axisymmetric solutions of higher-dimensional general
  relativity,''
  Phys.\ Rev.\  D {\bf 70} (2004) 124002
  [arXiv:hep-th/0408141].

\bibitem{Elvang:2005}
  H.~Elvang, R.~Emparan, D.~Mateos and H.~S.~Reall,
  ``Supersymmetric 4D rotating black holes from 5D black rings,''
  JHEP {\bf 0508} (2005) 042
  [arXiv:hep-th/0504125].

\bibitem{Ishihara:2006}
  H.~Ishihara, M.~Kimura, K.~Matsuno and S.~Tomizawa,
  ``Kaluza--Klein multi-black holes in five-dimensional Einstein--Maxwell
  theory,''
  Class.\ Quant.\ Grav.\  {\bf 23} (2006) 6919
  [arXiv:hep-th/0605030].

\bibitem{Evslin:2008}
  J.~Evslin,
  ``Geometric engineering 5d black holes with rod diagrams,''
  arXiv:0806.3389 [hep-th].

\bibitem{Ford:2007}
  J.~Ford, S.~Giusto, A.~Peet and A.~Saxena,
  ``Reduction without reduction: Adding KK-monopoles to five dimensional
  stationary axisymmetric solutions,''
  Class.\ Quant.\ Grav.\  {\bf 25} (2008) 075014
  [arXiv:0708.3823 [hep-th]].

\bibitem{Lu:2008}
  H.~Lu, J.~Mei and C.~N.~Pope,
  ``New black holes in five dimensions,''
  arXiv:0804.1152 [hep-th].

\bibitem{Belinsky:1971}
  V.~A.~Belinsky and V.~E.~Zakharov,
  ``Integration of the Einstein equations by the inverse scattering problem
  technique and the calculation of the exact soliton solutions,''
  Sov.\ Phys.\ JETP {\bf 48} (1978) 985
  [Zh.\ Eksp.\ Teor.\ Fiz.\  {\bf 75} (1978) 1953].

\bibitem{Belinsky:1979}
  V.~A.~Belinsky and V.~E.~Sakharov,
  ``Stationary gravitational solitons with axial symmetry,''
  Sov.\ Phys.\ JETP {\bf 50} (1979) 1
  [Zh.\ Eksp.\ Teor.\ Fiz.\  {\bf 77} (1979) 3].

\bibitem{Belinski:2001}
  V.~Belinski and E.~Verdaguer,
  ``Gravitational solitons,''
  Cambridge University Press, U.K.\ (2001).

\bibitem{Pomeransky:2005}
  A.~A.~Pomeransky,
  ``Complete integrability of higher-dimensional Einstein equations with
  additional symmetry, and rotating black holes,''
  Phys.\ Rev.\  D {\bf 73} (2006) 044004 
  [arXiv:hep-th/0507250].

\bibitem{Emparan:2004}
  R.~Emparan,
  ``Rotating circular strings, and infinite non-uniqueness of black rings,''
  JHEP {\bf 0403} (2004) 064 
  [arXiv:hep-th/0402149].

\bibitem{Iguchi:2006}
  H.~Iguchi and T.~Mishima,
  ``Solitonic generation of five-dimensional black ring solution,''
  Phys.\ Rev.\  D {\bf 73} (2006) 121501
  [arXiv:hep-th/0604050].

\bibitem{Tomizawa:2006}
  S.~Tomizawa and M.~Nozawa,
  ``Vacuum solutions of five-dimensional Einstein equations generated by
  inverse scattering method. II: Production of black ring solution,''
  Phys.\ Rev.\  D {\bf 73} (2006) 124034
  [arXiv:hep-th/0604067].

\bibitem{Elvang:2007}
  H.~Elvang and P.~Figueras,
  ``Black Saturn,''
  JHEP {\bf 0705} (2007) 050
  [arXiv:hep-th/0701035].

\bibitem{Emparan:2008}
  R.~Emparan and H.~S.~Reall,
  ``Black holes in higher dimensions,''
  arXiv:0801.3471 [hep-th].

\bibitem{Pomeransky:2006}
  A.~A.~Pomeransky and R.~A.~Sen'kov,
  ``Black ring with two angular momenta,''
  arXiv:hep-th/0612005.

\bibitem{Lu:2008b}
  H.~Lu, J.~Mei and C.~N.~Pope,
  ``New charged black holes in five dimensions,''
  arXiv:0806.2204 [hep-th].

\bibitem{Reall:2002}
  H.~S.~Reall,
  ``Higher dimensional black holes and supersymmetry,''
  Phys.\ Rev.\  D {\bf 68} (2003) 024024
  [Erratum-ibid.\  D {\bf 70} (2004) 089902]
  [arXiv:hep-th/0211290].

\end{thebibliography}
\end{document}